# E pur si muove!

*I.S.Nurgaliev*

*Department of Physics, Moscow State Timiryazev University,
Timiryazevskaya Str., 49. Moscow, 127550, Russia*

*E-mail: ildus58@mail.ru*

Abstract. The averaged term of squared vorticity is attributed to accelerated expansion. Cosmological singularity has been a consequence of excessive symmetry of flow, such as "Hubble's law". More realistic one is suggested. The new cosmological principle is applied to irregularities – they are homogeneous and isotropic in average to some extent within the corresponding Mega-galactic scales. The "Big Bang" is in fact the local bounce of Meta-galaxy which is typical among myriads of other bounces. Exact dynamic and steady/static solutions are presented. A "negative radiation" equation of state $p = \epsilon/3$ with $p \leq 0$, $\epsilon \leq 0$ is generated by vorticity which is the dynamic carrier of the dark energy. This fact dismisses the need in cosmological term, in any modifications of the gravity theory or in an exotic matter. A new concept of a material point is established. The paper comments as well on social and educational aspects of the findings.

Key words: cosmology, Universe, singularity, vorticity, bounce, nonsingular evolution, gravity, material point.

Нургалиев И.С. **E pur si muove!**

Аннотация. Утверждается, что завихренность и сдвиг, кинематические гидродинамические понятия, могли бы служить первыми простыми кандидатами на замену соответственно темной энергии и темной материи. Космологическая сингулярность оказывается следствием необоснованно высокой симметрии закона Хаббла. Выдвинута более реалистическая тензорная замена этого закона. Тем самым космологический принцип применяется не к распределению материи, а к распределению ее неоднородностей и их движения. Получается вывод, что «Большой Взрыв» несингулярен и он представляет собой локальный Мета-галактический отскок с ядерными реакциями, рядовой среди мириад других. Представлены динамические стационарные/статические космологические решения. Указано, что завихренностью, носителем темной энергии, генерируется эффективное уравнение состояния «негативного излучения» $p = \epsilon/3$ с $p \leq 0$, $\epsilon \leq 0$. Эти выводы делают ненужными пересмотра классических теорий гравитации и привлечения экзотических моделей материи. Возникает новая модель материальной точки второго рода, которая вращается и деформируется, позиционируясь по отношению к остальной части вселенной по Маху. Возникающая вращательная компонента скорости оказывается мнимым псевдовектором и перебрасывает мост к квантовой физике через концепцию классического спина и новых нелинейностей, позволяющих возникновение новых статических и стационарных состояний (частиц и структур) в первоначально бесспиновой динамике (пра)материи.

Ключевые слова: космология, Вселенная, сингулярность, завихренность, отскок, несингулярная эволюция, гравитация, материальная точка.

Academician Isaak Khalatnikov mentioned at the 13[th] Marcel Grossman Conference (http://www.icra.it/mg/mg13/) Lev Landau suggesting that something is too symmetric in the models yielding singularities, and that this problem is one of the three most important problems of modern physics. The aim of this report is to show that singularities are, indeed, consequences of such an overly "symmetrical approach" in building non-robust (i.e. without structural stability) toy models with singularities. Such models typically apply a synchronous system of reference and "Hubble's law", neglecting not-to-be-averaged-out quadratic terms of perturbations (specifically, differentially rotational velocities, vortexes).

Only by accounting the overlooked factors instead of Einstein's ad hoc introduction of a new entity, which was later declared by him as his "biggest blunder", can we correctly interpret accelerated cosmological expansion, as well as provide possibility of static solution. Current common perception of the observed accelerated expansion is that there is need either in modifying the General Relativity or discover new particles with unusual properties. Interestingly enough, both ways are possible depending on what kind of system of reference and corresponding interpretation are chosen, a decision which is usually made depending on the level of "geometrization". We limit ourselves to the smallest modifications: those easiest to accept for observations with existing methods and data processing techniques. For example, with leaf-like diagrams in the frame of the standard model neither will we change general relativity nor hypothesize new mystical particles predominating in the Universe. On the contrary, we bring a missed part of the motion into old theory. All we are doing is getting rid of a long-lived sophism: "let us call a small portion of matter a material point; a point does not spin, therefore nether does the material point." Sorry. Any portion of matter may rotate and may spin.

Therefore, to avoid the need for a long discussion of metaphysics (as well as the "imaginary" character of *'vortical'* motion) in this short paper, let us introduce a material point of the second type. It is characterized by mass density, rather than by mass as the point of the first type (particle); and it rotates around, generally speaking, its own axes i.e. it spins. These axes constitute a continuum. Anything material (including the material point) has an orientation towards other parts of the Universe. Machs principle is back. Getting rid of local rotation in building General Relativity brought the theory to a spin-free material world. The overlooked important level of freedom of the material point has to be restored and accounted for. We are trying to do so here. For mainstream direction of modifying basic gravity equations as f(R)-theory see [1] and many others. This direction also hits the target and results in averting the singularities. New developments in cosmological distance measurement techniques [2] will provide an observational test of the theoretical constructs offered.

***Results.*** Radius vectors of particles (ie, galaxies in a cosmological context) and their velocities are objects of the different vector spaces. Therefore projecting one into another takes affinor $H_{\alpha\beta}$:

$$V_\alpha = H_{\alpha\beta} R^\beta . \quad (1)$$

Let the Greek indexes be three-dimensional, and let us limit ourselves for now to a Newtonian description because it is essential to emphasize the fact that apparent cosmological singularities, kept cosmological thought tense, are not phantoms of special or general relativism but of the irrelevantly high level of symmetry of models of motion. We will escape Newtonian description when it is imperative and when we need to implant the result into observations based in General Relativity. The relationship (1) displaces the "Hubble law". The intuition of the great astronomer, who discovered a world of galaxies and who championed calculating measurements where there was a shortage of data, almost all his life kept him from fully accepting the too simplistic expression bearing his name. In 1931 Hubble wrote a letter to Willem De Sitter, one of the pioneers of new cosmology who predicted a new dynamic general relativistic effect that Hubble intended to measure, expressing his opinion on the theoretical interpretation of the redshift-distance relation: "... we use the term 'apparent velocities' in order to emphasize the empirical feature of the correlation. The interpretation, we feel, should be left to you and the very few others who are competent to discuss the matter with authority." (See in more details [3].)

His intuition did not fail. Transversal components (relative to the direction of observation, otherwise randomly oriented) of the cosmological motion ignored in the inventing Hubble law do contribute to cosmological redshift consistently with a Universe that is static-on-average-in-time globally and pulsating-waving locally, with an accelerated expansion phase in this pulsating-waving process (which is tremendously common in nature). Furthermore, as Robert P. Kirshner in [3] mentions -- with such epithets as "quaint and puzzling", "oddly, perhaps falsely, prescient" -- Hubble even made "allusion to acceleration" when everybody was expecting to see deceleration. His appeal to the theoreticians sounds like he feels that there could be a richer, more complex field to model turbulent motion; that there could be something else – perhaps a transversal redshift effect as we see here — to explain the redshift-distance correlation he endorsed.

The contributions of transversal peculiar motions to redshift are essentially nonlinear (quadratic for nearer galaxies, no linear approximation), while the longitude components contribute starting with linear term which, however, has the opposite sign of the famous Hubble constant contribution if its amount is seen as not averaged up to intergalactic cosmological scales: Andromeda is approaching the Solar system. These conclusions are clearly drawn from the expansion written in a special relativistic approximation. (The effect is necessarily relativistic, as the non-Doppler component does not have a Newtonian approximation) [4-7]:

$$\frac{v}{v_0} = 1 - RH/c + \frac{1}{2}(3H^2 - \omega^2)(RH/c)^2 - \frac{1}{2}H(11H^2 - \omega^2)(R\omega/c)^3 \quad (2)$$

of the nonlinear exact expression

$$\frac{v}{v_0} = \frac{\left[1 - \frac{R^2(\omega^2 + H^2)}{c^2}\right]^{1/2}}{1 + \frac{RH}{c}}. \quad (3)$$

Here vorticity $\omega$ and the Hubble parameter $H$ are components of the new homogenous and isotropic tensor

$$H_{ij} = \begin{pmatrix} H & \pm\omega+\sigma & \pm\omega+\sigma \\ \mp\omega+\sigma & H & \pm\omega+\sigma \\ \mp\omega+\sigma & \mp\omega+\sigma & H \end{pmatrix}. \quad (4)$$

R is the distance from the detector to an emitter (i.e., an atom of a star in a galaxy), $\sigma$ is shear. Note that even if we have $<\omega> = 0$, nevertheless $<\omega^2> \neq 0$ in (2) and (3). So Hubble might draw a parabola with "his" very few data starting from zero into the negative quadrant (blueshift), then, behind Andromeda, crossing the x-axis; then, after fitting for a while approximately to the linear "Hubble law", as any curve would, the demonstration acceleration would become clearer and clearer, again, and for a while quadratic; then it would show cubic correction as in (2) and so on fitting better and better to (3). The quadratic terms and thereafter, cubic terms, etc, easily derived from (3) (which, in general relativity, takes more careful consideration) are good landmarks to verify (or falsify as Popper put it) theories by observations. Plank's observations can provide this.

Differentiating the definition of the fundamental affinor (1) and using it one more time, we have the fundamental framework for differential laws of the mechanics encompassing the rotational part of the material point's motion of the multi-component media as well (the material point of the second type).

$$\dot{H}^b{}_{ij} + H^b{}_{ik}H^b{}_{kj} = \sum_a F^{ba}{}_{ij}(G,c,\hbar,e...). \quad (5)$$

WHS of (5) represents fields (tensions), gradients of forces numerated by *a*, acting on the matter component *b*, traditionally considered as causes of non-inertial accelerated motion. The second term in LHS includes kinematic factors sometimes associated with the so called 'vicious' forces (centrifugal, Coriolis). Note that $\omega = const(t) \neq 0$ and $\sigma = const(t) \neq 0$ are components of inertial motion. This is correct mechanics for continuous media. Spinning of material point is inertial motion. It has classic (non-quantized) origin. Velocity of the material point of the second type is presented by two vectors and one pseudo-vector (corresponding to diagonal, non-diagonal symmetric, and anti-symmetric parts of the affinor respectively). The generalization to arbitrary dimensional space is straightforward. The sum of the diagonal elements of the arbitrary tensor (which is not necessarily symmetric), separated from anti-symmetric and trace free elements is called *expansion* (meaning related speed of expansion) and in the presence of only Newtonian gravity (5) becomes for single component of matter

$$\dot{H}+H^2 = \omega^2 - \sigma^2 - \frac{4\pi G}{3}\rho, \qquad (6)$$

Where $3H = H_{11} + H_{22} + H_{33}$. Including vorticity and shear on the whs of nonsingular cosmologic equation (6) derived in Newtonian approximation is a tribute to Newtonian tradition and his second law, which includes here "new" cosmologic forces (inertial) additional to the "old" gravity, even though they are missing elements of the kinematics of old mechanics. Is the Coriolis factor an acceleration or a force? A pupil's answer before looking at (6) might depend on the books he/she has read. In geometry there is no such question though. Both the physical interpretations and precise terminology exist. In General Relativity we have settled on the interpretation of these missed elements of kinematics in cosmology not as "forces" but so-called "dark energy" and "dark matter". Earlier it took Einstein to invent the cosmological term to account for them in a phenomenological manner. To be specific, the concept of the cosmological term was in fact the introduction of vorticity as a counteragent to gravity, without specifying that and without stating that explicitly. Homogenous and isotropic vorticity is connected with the idea of the vorticity domain. If the standard Big Bang is correct back to the radiation-dominated period as the author believes for the most part, then we understand a lot of the "hot Universe" and can now even construct the static Universe, using Einstein's model, but without the cosmological term.

By taking H=0 in (6) for static universe, we have the cosmological solution $\omega^2 - \sigma^2 + \frac{4\pi G}{3}\rho = 0$. In the general relativistic description, we have the solution $\omega^2 - \sigma^2 + \frac{4\pi G}{3}(\rho + 3p) = 0$ which Einstein intended to derive [6]. As far as $\langle \omega^2 \rangle \neq 0$ necessarily, the model is probably better called quasi-static or steady state. But the latter name is taken and if re-used could cause confusion with the terminology. It can easily be shown that this static universe is stable against homogeneous and isotropic perturbations in contrast to the cosmologically constant, Einstein static Universe, as shown by Tolman [8, 9]. Therefore, the model based on the static solution can oscillate before getting destroyed by additional factors. One of the pulses is our bounce which we can see due to the redshift of galaxies. Discovery of cosmological background radiation shows that bounce was accompanied by nuclear reactions and close analogy between the Big Bang and zero-dimensional bomb explosion model exists. So, Meta-galaxy, as the local region of the Universe available for scientific exploration, is not static, even though the Universe, as we have seen, can be static on average, satisfying us esthetically and some of us (though not all of us) religiously twinkling with myriads of other meta- and mega-galaxies.

In deriving exact dynamic cosmological solutions, a good surprise was waiting for us. It is the same type of power-functional dependence of $\omega^2$ and $\sigma^2$ on $R$ as of the energy density and pressure of ultra-relativistic matter (electromagnetic radiation, "photon gas"). All of them, say, while isotropic, are proportional to $1/R^4$ and remain the same as the qualitative dynamic character of the cosmological equation (6) unchanged, causing only the re-defining of the constant $K^2 = \Omega^2 - \Gamma^2 - \Sigma^2$ as $(K')^2$ in the following equation (7). ( The following redefinition of $K$ is assumed: constants $\Omega, \Gamma, \Sigma$ stand for vortex, radiation (energy density and pressure) and shear constants in the corresponding conservation laws).

In Newtonian cosmology we derive in traditional notations:

$$\dot{H}+H^2 = K^2 \rho^{4/3} - (4\pi G/3)\rho. \quad (7)$$

Or, as an equation of the oscillator type with variable frequency

$$\ddot{R} = [(K^2 \rho^{4/3} - (4\pi G/3)\rho R)]R. \quad (7a)$$

The first integral of (7) holds $H^2/2 = -K^2/2R^4 + GM/R^3 + A/R^2$, where A is a constant of integration. A final integral of the cosmological equation follows: for $A \geq 0$ we have:

$t + t_0 = -(2A)^{-1}(2AR^2 K + 2GMR - K^2)^{1/2} - GM(2A)^{3/2}\ln(2^{3/2}A^{1/2}(2AR^2 + 2MGR - K^2)$
$+ 4AR + 2GM)$; for $A \leq 0$ we have $t + t_0 = (2A)^{-1}(2AR^2 K + 2GMR - K^2)^{1/2} - GM(-2A)^{3/2}$
$\arcsin[(2AR + GM)/(2A^2 K^2 + GM^2)^{1/2}]$.

We see that the Bing Bang was the Big Bounce Bang (BBB), the phase diagram of which is drawn in [5]. What is the exact value of $\omega^2$ in the radiation-dominated Einstein static Universe? We have

$$\omega^2 = \frac{4\pi G}{3}\rho_{radiation} = \frac{4\pi G}{3c^2}aT^4, \text{ where } a = \frac{\pi^2 k^4}{15\hbar^3 c^3} = 7.56 \times 10^{-16} J/m^3 K^4. \text{ Therefore, } \omega^2 = \frac{4\pi^3 G k^4}{45\hbar^3 c^5}T^4.$$

Or, compactly, $\omega^2 = a\chi T^4$, and $\omega^2 = \aleph T^4$, where $\chi = \frac{4\pi G}{3c^2} = 3.1 \times 10^{-27} m/kg$.

$\aleph = \frac{4\pi^3 G k^4}{45\hbar^3 c^5} = 2.11 \times 10^{-41} c^{-2} K^{-4}$ is a new universal constant combination. For T=2.7K we get that even such a small value as $\omega^2 \approx 1.12 \times 10^{-40}$ rad$^2$/c$^2$ could be enough to compensate the radiation contribution to the cosmological contraction which preceded the observed expansion, i.e. less than, supposedly, existing value. If, according to WMAP data, the density of "dark energy" is $9.9 \times 10^{-30}$ g/cm$^3$ then $\omega \approx 9.15 \times 10^{-20}$ sec$^{-1}$.

We conclude from dynamic equations that vorticity related dark energy is equivalent to "negative radiation" p=ϵ/3 with p≤0, ϵ≤0. This is a convenient way of presenting it for observational astrophysicists to see it first on their diagrams. Contribution of the shear (probably, dark matter) has the same effective equation of state but p≥0, ϵ≥0 if it comes from the same mechanism. There are arguments, though, holding that dark matter is composed of pressure-free heavy particles. At the same time, the argument for the author's hypothesis is the fact that this is the same effective equation of state that gravitational waves have. So, again it is not that important. The hypothesis leads re-defining another constant in the cosmologic equation, corresponding to non-relativistic matter. In addition, in contrast to vorticity $\omega$, shear σ can be removed by choosing the proper reference system.

***Discussion.*** We can compare this result with Einsten's approach. In the Newtonian framework he assumed first

$$\Delta\varphi + \lambda\varphi = 4\pi G\rho.$$

We see why Einstein's attempts [8] failed. The correct ansatz is

$$\Delta\varphi + \omega^2 - \sigma^2 = 4\pi G\rho.$$

Going further, Einstein might arrive using a general relativistic description to the conclusion

$$G_{\mu\nu} + \lambda g_{\mu\nu} = \kappa T_{\mu\nu} \Rightarrow \lambda = -2\omega^2 + 2\sigma^2.$$

But instead of the traditional product of the cosmological term and the metric tensor, our approach brings us to an additional hydrodynamic energy-momentum tensor for cosmology which has a very familiar looking but in fact "exotic" effective equation of state of "negative radiation" standing for dark energy: p =ϵ/3 with p≤0, ϵ≤0. It may have another possible term with positive energy of "dark matter" standing for energy of shear motion p =ϵ/3 with p≥0, ϵ≥0. Whether the product in fact has this could be tested by observation. Therefore, the name "cosmological term" in this paper does not mean constant scalar lambda multiplied by the metric tensor but rather a fragment of the complex energy-momentum tensor generated by factors σ and ω missed earlier and brought into cosmology now. Neither the cosmological term nor dark energy is needed. *"Hypotheses non fingo" de facto.*

The Copernican principle develops into a perfect cosmological principle through the presented solutions, which, again, are locally dynamic, Meta-galactic, and cosmologically-globally average-in-time static (meaning averaged over cosmological periods, circles, bounces): neither our location in space nor the period of humankind's flourishing is special. We live during one of the pulses. The pulses are not singular; on the contrary, bounces are common phenomena in the Universe. The "bang" is "big" in comparison with non-cosmological "bangs", but just ordinary compared to other cosmological ones, except the fact that, yes, we live in this one. The current phase of the pulse has acceleration. Every cyclical process has an acceleration period after its bounce. Our pulse was deep; it included nucleosynthesis during the bounce, and cosmological soup [11] was cooked, its expansion phase started from the bounce on the effective potential of the Meta-galactic vorticity.

As to the famous Penrose-Hawking theorems [12-13], they, in fact, can be read in the given context as statements that we do not know other factors in the framework of standard General Relativity and the usual classic models of matter that are able to resist forming trap surfaces except vorticity. Look at the sky or at photos of the galaxies taken with the Hubble telescope. Is not the cosmos full of vortexes, spinning? Some do not see even the Sun revolving around the Earth, many call it apparent. Religious tradition teaches trust in authorities, not to your eyes. In fact, everything revolves around everything else and spins. So does the material point of the second type. So does the Meta-galaxy as part of the Universe. "E pur si muove!"

It should be noted that solutions with vorticity are robust (in contrast to singular ones) as it should be for models of reality. This enables us to try to build on the realistic model other aspects of the evolution scenario including nonlinear factors such as clustering, nuclear reactions. "Initial" conditions of cosmological perturbations along with the "birth of the Universe" are left for metaphysical inquiry.

The rates of cosmological and meta-galactic processes are very slow during the long period of their lifespan in comparison to the lifespan of the objects (e.g., stars, galaxies). This peaceful cosmological scenario is studded with sudden, catastrophic events involving the explosion of the cosmic objects. This fact is typical for the nonlinear dynamics of at least the quadratic (and higher) type. The Big Bounce Bang of the Meta-galaxy is the biggest one we are witnessing directly inside in this eternal Universe.

*Methods.* The method used in this article is nonlinear mathematical modeling. The concept of the nonlinear reaction-advection-diffusion equation along with the second type material point harmoniously provides a bridge between classical and quantum physics and explains why matter is organized discretely with different levels of organization (from clusters of galaxies to atoms and elementary particles). We may expect that system of evolutionary equations

$$\frac{\partial \rho_i}{\partial t} = -\tfrac{1}{3} H_i \rho_i + f(\{\rho_i\}) + \nabla\left[D_i(\{\rho_i\}) \nabla \rho_i\right]$$

describes arbitrary amount of material components with densities $\rho_i$, Hubble parameters $H_i$ and coefficients of the effective diffusion $D_i$, generalized and adopted when needed. It may also provide nonlinear evolution scenarios for evolution of the cosmological perturbations in the multi-component cosmologic reacting media. The nonlinear term $f(\{\rho_i\})$ stands for mutation of the components (i.e., nuclear and chemical reactions, gravitational clustering, and cosmologic morphogenesis). A linear consideration of the cosmological perturbations within demonstrated method for the ***arbitrary amount*** of components done in [14] in the form of exact analytical solutions in terms of higher transcendental G-functions of Maijer. This method in the frame Einstein-Cartan theory gave static cosmological solution [15]. The last remark is: so, sometimes, more complicated theories help understand simpler ones.

Acknowledgments

The author is grateful to Academicians A.A.Starobinsky, R.A.Synyayev, I.M.Khalatnikov as well as to G.S.Islamov for simulative discussions.